\documentclass   [a4paper,epsfig,12pt]{article}
\usepackage{graphicx}
\begin{document}
\begin{flushright}
               FZU-D 20020206
\end{flushright}

\begin{center}
   {\bf Quantum mechanics and EPR paradox } \\[3mm]
    Milo\v{s} V. Lokaj\'{\i}\v{c}ek \\
     Institute of Physics, AVCR, 18221  Prague 8, Czech Republic \\
\end{center}

Abstract

The orthodox quantum mechanics has been commonly regarded as being
supported decisively by the polarization EPR experiments, in which
Bell's inequalities have been violated. The given conclusion has
been based, however, on several mistakes that have not been yet
commonly known and sufficiently analyzed. The whole problem will
be newly discussed and a corresponding solution will be proposed.
\\

 {\bf 1. Introduction  }

It is commonly believed that the Copenhagen interpretation of
quantum mechanical model represents the only possibility of
describing physical processes of microworld. The main support for
such statement is seen in the results of experiments proposed in
principle by Einstein, Podolsky and Rosen in 1935 \cite{ein}. The
way to their interpretation in agreement with Copenhagen school
has been, however, paved at least by three mistakes.

The first one coming from von Neumann \cite{neu} in 1932 has been
step-by-step discovered already earlier (G. Hermann in 1935
\cite{herr}, D. Bohm in 1952 \cite{bohm}, J. Bell in 1964
\cite{bell}). However, the other two mistakes have not been known
and discussed to a sufficient extent until now. One of them
relates to Bell's inequalities as their derivation has been based
on one assumption, the impact of which has not been generally
recognized; the assumption being hardly acceptable for the common
type of polarization EPR experiments. It has not been possible to
derive these inequalities without such an assumption. The
corresponding analysis of the problem may be found in Ref.
\cite{loka}. As to the third mistake (see the book of Belifante
\cite{belif}) it has had probably yet more important impact on the
conviction of the most physicists. F. Belifante argued in 1973
that practically any hidden-variable theory was to provide
significantly different predictions from those derived with the
help of standard quantum-mechanical model, which is not true. It
will be shown in the following that already a very simple
hidden-variable theory gives practically the same predictions for
a pair of polarizers (including EPR coincidence experiments) as
the standard quantum mechanics.

Some important differences between hidden-variable interpretation
and quantum mechanics should exist, of course, e.g., in the light
transmission through three polarizers. Such experiments inspired
by preliminary theoretical results were  performed and published
in 1993 and 1994 (see \cite{kra1,kra2}); they have shown that the
standard quantum-mechanical theory of polarized light should be
regarded in principle as falsified by these experimental results.
A way of interpreting these experimental results on a new basis
should be looked for.

The problem concerning the derivation of Bell's inequalities will
be summarized shortly in Sec. 2. The misleading argument of
Belifante will be discussed in Sec. 3; a very simple
hidden-variable model of light transmission through a polarizer
pair will be described. A generalized model taking into account a
polarization shift (towards the polarizer axis) during the light
passage though a polarizer will be then proposed in Sec. 4.
 The transmission characteristics (for individual polarizers)
derived to be in agreement with the Malus law will be then used in
predicting transmission characteristics for a triple of polarizers
(Sec. 5); the results will be compared to quantum-mechanical
predictions. The consequences following from the experimental data
obtained with three polarizers will be discussed in Sec. 6.
\\

 {\bf  2.  Light transmission through a polarizer }

The transmission of light through a polarizer has been studied
since the beginning of  the 19th century. It was found that the
intensity of unpolarized light passing through two polarizers was
decreasing according to Malus law, i.e. as
\begin{equation}
 m(\alpha)  \;= \; (1-\varepsilon)cos^2\alpha \;+\; \varepsilon  \label{gemal}
\end{equation}
where  $\alpha$ was the angle between polarizer axes. It was
assumed that $\varepsilon = 0$ at least for the so called ideal
polarizers; in any case $\varepsilon\,\ll\,1$.

Einstein argued in the thirtieths years that the quantum mechanics
was not a complete theory and that some more detailed
characteristics were necessary to be added to describe fully a
microscopic object. However, physical community had not accepted
his critical point of view.  The main support for standard quantum
mechanics was seen at that time undoubtedly in the "proof" of von
Neumann that any "hidden" variables were excluded by the
quantum-mechanical model. It was not taken into account, either,
that already in 1935 Grete Herrmann \cite{herr} showed that the
approach of von Neumann was practically a "circle proof". The
argument of D. Bohm \cite{bohm} that a hidden variable was
contained already in Schr\"{o}dinger equation was accepted
seriously by a very small number of the then physicists. Only the
approach of J. Bell \cite{bell} met with greater attention,
especially since formulas were presented that seemed to enable
bringing a decision between the two (orthodox and ensemble)
interpretations of the quantum-mechanical model on experimental
basis.

However, in a broad physical community there has not been any
interest to change a generally accepted paradigm. Any greater
doubts about the standard theory have not been evoked from the
fact, either, that it was navigated from the very beginning by the
mistake of von Neumann. The firm belief in the so called EPR
paradoxes seems to live still in a great part of physical
community. The main reason may be seen in that they have been
supported seemingly by other two already mentioned arguments, both
being false.

The first of these two arguments has related (as already
mentioned) to Bell's inequalities that have been believed to hold
for any hidden-variable alternative. However, their application to
the current coincidence polarization experiments cannot be
regarded as regular. In their derivation a seemingly self-evident
assumption has been made use of. To derive these inequalities it
has been necessary to interchange always the transmission
probabilities for photons belonging to different photon pairs,
which has been equivalent to assuming for transition probabilities
of individual photons to be independent of the impact point into
the internal (plane grid) polarizer structure; or to regarding the
corresponding measuring devices in principle as at least
half-black boxes. Detailed analysis of the problem may be found in
\cite{loka}.

Bell's inequalities cannot be derived without the given
assumption. Consequently, the violation of Bell's inequalities in
the common EPR experiments does not provide any argument against a
hidden-variable alternative, as they do not correspond to a fully
consistent hidden-variable (realistic) description.

However, as already mentioned there has been another mistake more
that has had probably a yet more important impact in influencing
the attitude of physical community towards the belief in EPR
paradoxes, which will be discussed in the next section.
\\

 {\bf  3. Malus law and photon transmission through one polarizer }

  Belifante argued in his book \cite{belif} that the standard
quantum mechanics and a hidden-variable theory should lead to
quite different predictions as to current EPR experiments.
However, such a statement has not been true, which will be now
demonstrated. The angle dependence of light transmission through a
polarizer pair in  a hidden-variable theory may be expressed as
\begin{equation}
       p_2(\alpha)  =   \int^{\pi/2}_{-\pi/2} {\bar p}_1(\lambda,0)
             \;  {\bar p}_2(\lambda,\alpha) \; d\lambda
\end{equation}
where $\alpha$ is again the angle between the axes of polarizers
and $\lambda$ is photon polarization; ${\bar p}_j(\lambda,\alpha)$
being transmission probability of a photon characterized by
$\lambda$ polarization through a polarizer deviated by angle
$\alpha$ from the same zero direction. Belifante has chosen quite
arbitrarily
\begin{equation}
     {\bar p}_j(\lambda,\alpha) \;=\; p_1(\lambda-\alpha)
                             \;\sim\; \cos^2(\lambda-\alpha) \;,
\end{equation}
which has led to fundamental deviations of $p_2(\alpha)$ from the
Malus law (\ref{gemal}).

However, the problem should have to be solved in opposite way. The
question has been, which function  $p_1(\lambda)$ corresponds to
the Malus law. The actual solution of the problem is represented
by the full line in Fig. 1. The given curve may be described,
e.g., by the formula
\begin{equation}   \label{pfi}
         p_1(\lambda-\alpha) \;=\; [1-\phi(|\lambda-\alpha)|]
\end{equation}
where
\begin{equation}
  \phi(\gamma) \;=\;  [1-\exp(-a\gamma^{e})]/[(1 + c\exp(-a\gamma^{e})] \;;
     \;\;\;a,e,c \,>\,0 .
                      \label{phi}
\end{equation}

 The  transmission probability represented by the full line in Fig. 1
is given by the following values of free parameters in Eq.
(\ref{phi}):
   \[ a \;=\; 1.74 , \;\;\; e \;=\; 3.78 , \;\;\; c \;=\; 200 . \]
The light distribution around the axis outgoing from the first
polalizer  is characterized by $d(\lambda)$; comp. similar
function obtained under more general conditions and shown in Fig.
2.

The dependence $p_2(\alpha)$ of light transmission through a
polarizer pair on mutual angle values corresponds to the
generalized Malus law for higher values of $\alpha$ very well; see
Fig. 1. As to the deviations at smaller angles there have not been
suitable data for a detailed comparison; it would be very
interesting to perform a thorough comparison of theoretical
predictions in the whole angle range. In any way, we must conclude
that there is not surely any significant difference in predictions
for available EPR coincidence measurements; Belifante's graph
\cite{belif} must be denoted as false.

A better agreement with Malus law may be obtained with the help of
probability function  $\,p_1(\lambda)\,$  represented by a greater
number of free parameters. However, even in such a case some
greater deviations remain in the region of small angles, which
might indicate that or the Malus law is not fully exact at the
given angles or some mechanism exists that changes polarization of
a photon passing through a polarizer. The latter possibility will
be followed in Sec. 4.    \\

  {\bf  4.  Generalized transmission model }

In Sec. 3 we have assumed that the spin or polarization of a
photon does not change its direction in passing through a
polarizer. However, some data seem to indicate that polarization
may shrink to polarizer axis more than given by mere transmission
probabilities. Let us consider now such a possibility.

We will assume that $\lambda$-distribution of photons outgoing
from the first polarizer is not given by the function
$\;d(\lambda)\,=\,p_1(\lambda)/(\pi/2)\;$ (see Fig. 1), but that
it is given by
\begin{equation}
 d(\lambda) \;=\; \int_{-\pi/2}^{\pi/2} p_1(\lambda') c(\lambda,\lambda')d\lambda'
\end{equation}
where
\begin{equation}
      \int_{-\pi/2}^{\pi/2} c(\lambda,\lambda')d\lambda  \;=\; 1 \; ; \label{norm}
\end{equation}
i.e. that a photon having had  original polarization $\lambda'$
has gained polarization $\lambda$ with the probability
$c(\lambda,\lambda')\geq 0$. It holds then
\begin{equation}
       p_2(\alpha)  =   \int^{\pi/2}_{-\pi/2} d(\lambda)
             \;   p_1(\lambda-\alpha) \; d\lambda
\end{equation}

We have chosen the following parameterization for probability
function:
\begin{equation}
     c(\lambda,\lambda') \;=\; A_\sigma(\lambda') e^{-\sigma(\lambda'-\lambda_e)^2}
\end{equation}
where  $A_\sigma(\lambda')$  is normalization coefficient
(guaranteeing the validity of Eq. (\ref{norm})) and
\begin{equation}
 \lambda_e \;=\; \lambda [1 + \epsilon(\eta-\lambda)]\, ,
                   \;\;\; 0 < \lambda \leq \eta \,,      \label{lae}
\end{equation}
\begin{equation}
 \lambda_e \;=\; \frac{\pi}{2}-(\frac{\pi}{2}-\lambda)[1 +\epsilon(\lambda-\eta)]\,,
                   \;\;\;  \eta < \lambda < \pi/2 \; ;      \label{lae}
\end{equation}
 $\sigma$, $\epsilon$ and $\eta$ are free parameters;
$\sim\pi/4\, < \,\eta\,<\,\pi/2$. We have assumed that a greater
part of $\lambda$-polarizations shrinks towards the polarizer axis
while a rest keeps around a perpendicular direction.

The fit obtained under such conditions is shown in Fig. 2. The
given results corresponds to the following values of free
parameters
 \[ \sigma \; =\; 40.5, \;\;\;\; \epsilon\;=\; 0.40,  \;\;\;\; \eta \;=\; 1.38,  \]
    \[ a \;=\; 2.38 , \;\;\; e \;=\; 2.54 , \;\;\; c \;=\; 186.8 . \]
It holds also for total light transmissions
  \[ I_1/I_0 \,=\, \int^{\pi/2}_{-\pi/2}d(\lambda)d\lambda \,=\, 0.496, \;\;\;\;
       I_ 2/I_0 \,=\, \int^{\pi/2}_{-\pi/2}d(\lambda)d\lambda \,=\, 0.482\, . \]
\\

  {\bf  5.  Light transmission through a triple of polarizers  }

Both the theories (quantum mechanics and hidden-variable theory)
give practically the same predictions for light transmission
through a pair of polarizers. However, one must expect that these
predictions may significantly differ in other experiments, e.g.,
for the transmission of light through three polarizers. To analyze
this case we will return to the simpler version introduced in Sec.
3 (without a polarization change during the passage) and try to
derive corresponding characteristics of light transmitted through
three polarizers.
   In such a case it is possible to write (in hidden-variable alternative)
\begin{equation}
  I(\alpha,\beta)  \;=\;    \int^{\pi/2}_{-\pi/2} {\bar p}_1(\lambda)\;
           {\bar p}_2(\lambda-\alpha)\; {\bar p}_3(\lambda-\beta)\; d\lambda
\end{equation}
where $\alpha$ and $\beta$ are angle deviations of the second and
third polarizers to the axis of the first polarizer. We will
 assume that the transmission probabilities
 ${\bar p}_i(\lambda)=p_1(\lambda)$ are characterized by parameters derived
in Sec. 3. The polarization shrinkage during light passage through
a polarizer will be neglected.

According to standard quantum mechanics holding for ideal
polarizers (or to electromagnetic light theory) it should hold
\begin{equation}
          I(\alpha,\beta)  \;=\;   \cos^2\alpha\: . \:cos^2(\alpha-\beta) .
\end{equation}

The comparison of predictions by both the theories has been given
in Fig. 3. The dependence of light intensity on $\alpha\,$ (for
$\beta = 0\,$) indicates that measurable differences should exist
surely around $\alpha \;\simeq\; 50-75^o$. It means that a
decision between these two theoretical alternatives might be given
on experimental grounds. There is not any doubt that one should
come to reliable conclusions concerning the validity of individual
theoretical alternatives when the experimental measurement is
performed with the polarizers exhibiting very small value of
parameter $\varepsilon$ in Eq. (\ref{gemal}); being near to the so
called ideal polarizers.  \\

 {\bf  6.  Experimental data with three polarizers  }

The experimental data enabling to compare the results derived in
the preceding section have been published already earlier (see
Refs. \cite{kra1,kra2}). Some consequences of these results have
been mentioned in Ref. \cite{haif}. The experimental results do
not seem, however, to correspond well  to any of the given
predictions given in Sec. 5. Anyway, they are in a strong
disagreement with quantum-mechanical predictions, which should be
a challenge of looking for a new theoretical explanation of
polarization phenomena.

A full agreement has not been obtained with the results of Mueller
calculus based on the old phenomenological theory proposed by
Stokes, either, even if the data and predictions have exhibited
some similar features (see Ref. \cite{kra2}). A more detailed
analysis of the given experimental data will be  given elsewhere
later.     \\

  {\bf  7.  Conclusion  }

One can conclude that some mistakes have influenced the way to the
contemporary theory of microscopic physical world, which concerns
also the description of polarization phenomena. Having removed
these mistakes one is forced to look for a better description,
especially, of data concerning experimental results with different
numbers of polarizers; with the help of a suitable kind of
hidden-variable models.

It has been shown in Sec. 3 that a good approximate description
may be obtained already with a very simple model; it has been
assumed that photon passing through a polarizer does not change
its polarization (or direction of its spin). A much better
agreement with generalized Malus law may be obtained if some
shrinkage of polarization to polarizer axes occurs during the
light passage through a polarizer. Anyway, hidden-variable
alternative seems to open good possibilities of explaining all
experimentally established polarization phenomena.

\begin{figure}[htb]
\begin{center}
\includegraphics*[scale=.4, angle= -90]{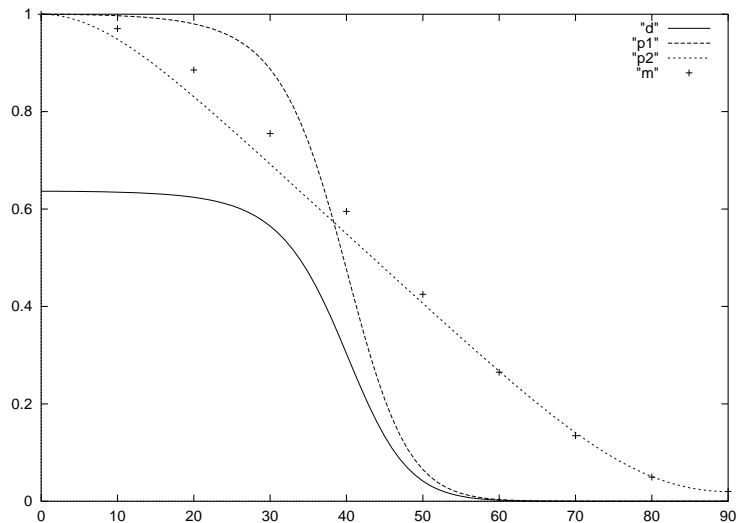}
   \caption{ {\it
Transmission probability through a polarizer pair leading to Malus
law:  $p_1(\lambda)$ - dashed line;  $p_2(\lambda)$ - dotted line;
$d(\lambda)$ - full line;  Malus law $m(\lambda)$ - individual
points.   } } \label{fg1}
\end{center}
 \end{figure}
\vspace{0.2cm}

\vspace{1cm}
\begin{figure}[htb]
\begin{center}
\includegraphics*[scale=.4, angle= -90]{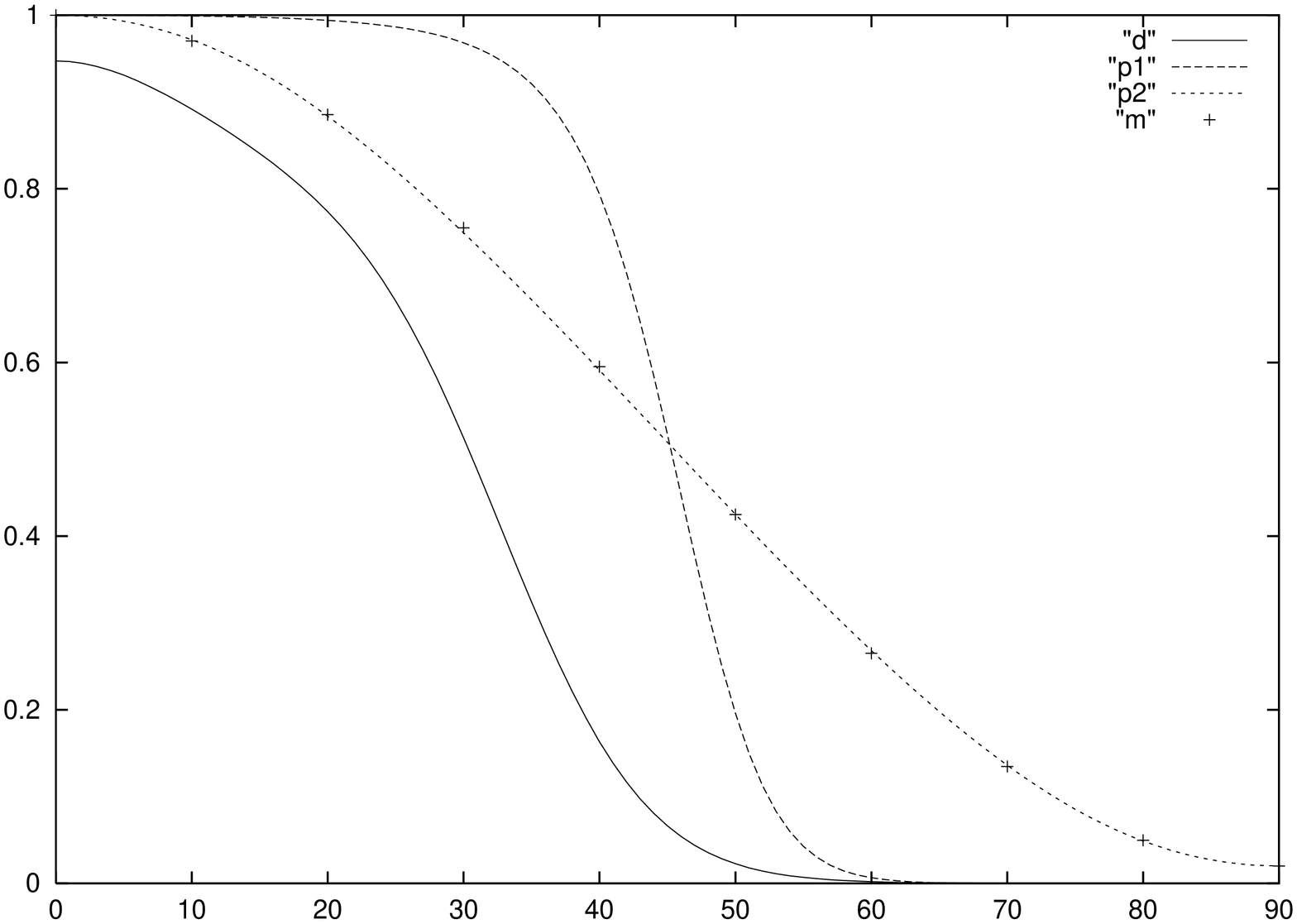}
   \caption{ {\it
     Extended model of transmission probability through a polarizer pair
             leading to Malus law:  $p_1(\lambda)$ - dashed line;
     $p_2(\lambda)$ - dotted line; $d(\lambda)$ - full line;
     Malus law $m(\lambda)$ - individual points. }}   \label{fg2}
\end{center}
 \end{figure}
\vspace{0.2cm}

\vspace{1cm}
  \begin{figure}[htb]
\begin{center}
\includegraphics*[scale=.4, angle= -90]{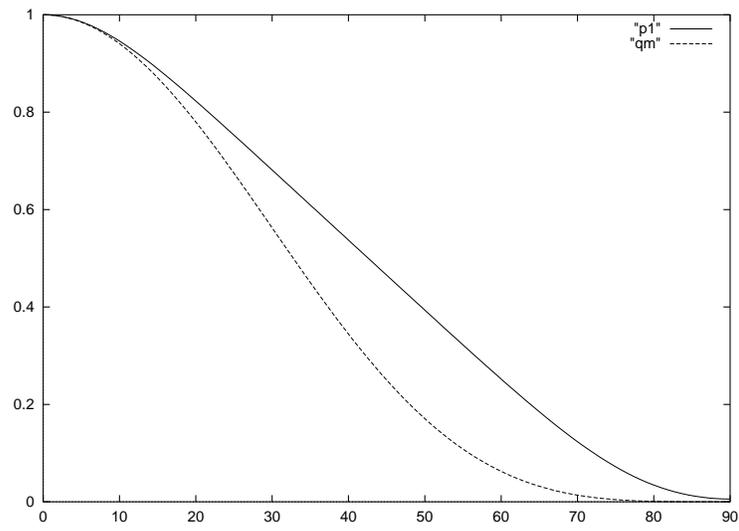}
   \caption{ {\it    Light transmission through three polarizers (dependence on $\alpha$
            at $\beta = 0$);  full line - hidden-variable alternative,
            dashed line -  quantum mechanics.
   } } \label{fg3}
\end{center}
 \end{figure}
\vspace{0.2cm}

\end{document}